\documentclass[aps,prl,epsf,epsfig,floatfix,groupedaddres,superscriptaddress,twocolumn]{revtex4-1}
\usepackage{graphicx}
\usepackage[usenames,dvipsnames]{xcolor}
\usepackage{amsfonts}
\usepackage{amsmath}
\usepackage{amssymb}
\usepackage{verbatim}
\usepackage{simplewick}
\usepackage{eucal}
\usepackage{enumerate}
\usepackage{subfigure}
\usepackage{url}
\definecolor{uc}{rgb}{0,0.6,0.1}
\usepackage[breaklinks]{hyperref} 
\hypersetup{
     colorlinks=true,
     citecolor=blue,
     urlcolor=orange,
     linkcolor=red
}
\def\bkcolor{\textcolor{OrangeRed}}
\usepackage{mycommand}
\usepackage{natbib}
\providecommand{\affiliation}[1]{{\it{\centering #1}}\\}
\providecommand{\email}[1]{\footnote{#1}}
\input{jrnl.in}
\providecommand{\etal}{{\it et al.}}
\begin{document}
%
\def\zpi{\Pi^0}
\def\tpi{\Pi}
\def\stot{\Sig^{\text{tot}} }
\def\stotu{\Sig^{\mr{tot}}_\ua}
\def\stotd{\Sig^{\mr{tot}}_\da}

%

\author{Himadri Barman}
\title{Zero temperature dynamics of the Hubbard model in infinite dimensions: A local moment approach}
\email{hbarhbar@gmail.com}
\affiliation{Department of Theoretical Physics, Tata Institute of Fundamental Research, Homi Bhabha Road, Navy Nagar, Mumbai 400005, India}
%

\begin{abstract}
The local moment approach (LMA) has presented itself as a powerful semi-analytical quantum impurity solver (QIS) in the context of the dynamical mean-field theory (DMFT) for the periodic Anderson model and it correctly captures the low energy Kondo scale for the single impurity model, having excellent agreement with the Bethe ansatz and numerical renormalization group results. However, the most common correlated lattice model, the Hubbard model, has not been explored well within the LMA+DMFT framework beyond the insulating phase. Here in our work, within the framework we attempt to complete the phase diagram of the single band Hubbard model at zero temperature. Our formalism is generic to any particle filling and can be extended to finite temperature. We contrast our results with another QIS, namely the iterated perturbation theory (IPT) and show that the second spectral moment sum-rule improves better as the Hubbard interaction strength grows stronger in LMA, whereas it severely breaks down after the Mott transition in IPT. 
We also show that, in the metallic phase, the low-energy scaling of the spectral density leads to universality which extends to infinite frequency range at infinite correlation strength (strong-coupling). At large interaction strength, the off half-filling spectral density forms a pseudogap near the Fermi level and filling-controlled Mott transition occurs as one approaches the half-filling. Finally we study optical properties and find universal features such as absorption peak position governed by the low-energy scale and a doping independent crossing point, often dubbed as the \emph{isosbestic point} in experiments.   
\end{abstract}
\maketitle

\section{Introduction}
The Hubbard model (HM)~\cite{hubbard:prsa63} is the simplest model that incorporates on-site
correlation effect between electrons in a lattice.
 Despite its appealing simplicity  and applicability
 to Mott metal-to-insulator transition~\cite{mcwhan:rice:remeika:prl69}, 
and high-temperature superconductivity~\cite{bednorz:muller:zpb86},
 the model has remained
a daunting challenge to the condensed matter physicists. It
has been studied extensively from many angles including mean-field analytics
and exact diagonalization numerics~\cite{penn:pr66,hirsch:prb85}. 
During the past decades,  the dynamical mean-field theory
(DMFT) has exhibited itself as an extremely powerful numerical method 
that simplifies the lattice model problem by mapping onto an effective interacting single impurity problem self-consistently connected to a fermionic bath via 
a hybridization function~\cite{georges:kotliar:krauth:rozenberg:rmp96}. 
The mapping becomes exact at infinite coordination number
of the lattice and the self-energy becomes moment-independent at that limit.
Even after the advent of the DMFT, solving an interacting lattice model
remained elusive at the level of the effective impurity model problem.
Therefore apart from the challenges that arise due to additional complication of 
a model (e.g. multiple orbitals, spin-orbit interaction, electron-phonon coupling etc.),
finding a suitable quantum impurity solver (QIS) for DMFT method is still 
an ongoing issue. In addition to this, a quick or computationally less 
expensive QIS is required for systems having multiple bands, multiple layered structures,
finite cluster sizes, and consisting of other real material-based parameters. 
In fact, besides the holy grail of getting the most accurate QIS, a race 
is going on towards achieving the fastest QIS, which can at least 
capture the qualitatively correct physics and energy scales associated to 
it~\cite{zhuang:wang:fang:dai:prb09,feng:zhang:jeschke:prb09}.

Depending on the invention of several QISs we may divide 
DMFT timeline into two major decades starting from early nineties.  
At the first decade several methods came up as candidates
of the QIS, such as the iterated perturbation theory (IPT)~\cite{georges:kotliar:prb92,georges:krauth:prb93,zhang:rozenberg:kotliar:prb93,rozenberg:kotliar:zhang:prb94}, exact diagonalization (ED)~\cite{caffarel:krauth:prl94}, Hirsch-Fye quantum Monte Carlo (HFQMC)~\cite{jarrell:prl92,georges:krauth:prl92,zhang:rozenberg:kotliar:prb93}, non-crossing approximation (NCA)~\cite{pruschke:cox:jarrell:prb93}, and numerical renormalization group(NRG)~\cite{bulla:hewson:pruschke:jpcm98,bulla:prl99}. All these methods could successfully capture the Mott metal-to-insulator transition by opening a gap at the Fermi level of the spectral density. However, all of 
them suffer from limitations. For instance, ED exhausts the computational limit before one achieves a reasonable size of a lattice;  HFQMC becomes disadvantageous at low temperature and suffers from the fermion sign problem~\cite{loh:etal:prb90}; NCA is only reliable for the insulating solution as the metal fails to promise a Fermi liquid; NRG becomes less accurate towards the high-energy (Hubbard bands) regime. Moreover, both ED and NRG suffer from energy discretization 
artifacts~\cite{zitko:pruschke:prb09}.

In the next decade, dynamical renormalization group 
(DMRG)~\cite{garcia:hallberg:rozenberg:prl04, nishimoto:gebhard:jeckelmann:jpcm04,
karski:raas:uhrig:prb08}, fluctuation exchange approximation (FLEX)~\cite{drachl:etal:jpcm05}, and comparatively more recently the  continuous time quantum Monte Carlo (CTQMC)~\cite{rubtsov:savkin:lichtenstein:prl05, werner:comanac:medici:troyer:millis:prl06, park:haule:kotliar:prl08, gull:millis:lichtenstein:rubtsov:troyer:werner:rmp11} came up.
FLEX becomes limited to a certain range of interaction 
strength~\cite{kotliar:etal:rmp06}. 
On the other hand, though CTQMC can promise to work at very low temperature, it requires analytical continuation in order to get physical quantities in real frequency and the method of doing so is tedious and introduces additional errors~\cite{jarrell:gubernatis:prp96}. Moreover, ED, DMRG and CTQMC methods all demand expensive computational challenges.

Therefore if we just want to seek a semi-analytical method, apart from the IPT and the NCA, another QIS, namely the local moment approach (LMA) deserves an attention. LMA was pioneered by Logan and his co-workers in the end of the first decade and it became very efficient in capturing 
the low energy Kondo scale in the single-impurity 
Anderson model (SIAM), and its strong-coupling behavior (infinite Hubbard interaction) 
shows excellent agreement with Bethe ansatz~\cite{logan:eastwood:tusch:jpcm98} and NRG results~\cite{dickens:logan:jpcm01}. Within DMFT framework, LMA has been extensively applied to the particle-hole symmetric and asymmetric periodic Anderson model (PAM) that corresponds to Kondo insulators and heavy fermionic systems respectively. In both cases the strong coupling (Kondo lattice limit) behavior of the low-energy scale has been captured well, and additionally the finite temperature transport and optical properties can explain many universal features found in the experiment~\cite{victoria:logan:hrk:epjb03,nsv:victoria:logan:jpcm03,nsv:logan:epjb04,nsv:logan:jpcm05,logan:nsv:jpcm05}. A recent study has exhibited how doping leads to mix valence to Kondo lattice crossover, in accord with such signatures found in transport and optical properties of several heavy fermion compounds~\cite{pramod:nsv:jpcm11}. In spite of all these successes, the Hubbard model,
has received less attention from the LMA aspect. 
Only the results of half-filling HM at large interaction have been reported, where LMA finds insulating spectral density for both the paramagnetic and antiferromagntic cases, and the strong-coupling Heisenberg ($t$-$J$) limit is captured correctly~\cite{logan:eastwood:tusch:prl96,logan:eastwood:tusch:jpcm97}. 

Here we extend the scenario for all interaction strengths and fillings. Recently a generic version of LMA with variational method~\cite{kauch:byczuk:pb12} was proposed for the multi-orbital
extension of the LMA. However, the method deviates from the conventional formalism, as already applied to SIAM and PAM, and does not ensure the Luttinger pinning (to be discussed in the forthcoming section) of the spectral density.
In principle,  our method could be applied to finite temperature as well, however, we restrict ourselves to the ground state only, leaving the finite temperature results a topic for a subsequent paper. We must mention, another important concern of modern day's QIS, is the obedience of sum-rules~\cite{turkowski:freericks:prb06, zitko:pruschke:prb09, ruegg:gull:fiete:millis:prb13, lu:hoppner:gunnarson:haverkort:prb14}, e.g. whether
the spectral moments from the numerics become closer to their exact (details are discussed in the results section). We discuss this aspect in LMA case and show that stronger the interaction, the spectral moment becomes more accurate.

Our work is organized as follows. We first describe the formalism for the half-filling, i.e. particle-hole (p-h) symmetric case, then we discuss the modification over it to deal the asymmetric case where we introduce an asymmetry parameter $\eta$. Then we show the numerical results, viz. spectral densities and properties derived from them and finally conclude. In places where required, we compare our results with another semi-analytical QIS, namely the IPT.

\section{Formalism}
As a part of formal introduction and for future references in the discussion part, 
we first write down the single band Hubbard model Hamiltonian below:
\blgn
\h H=-\sum_{\lngl i j\rngl,\sig} t_{ij} c\y_{i\sig} c\py_{j\sig}
+(\eps_d-\mu)\sum_{i\sig} c\y_{i\sig} c\py_{i\sig}
+U\sum_i \n_{i\ua} \n_{i\da}
\label{eq:HM}
\elgn
where $t_{ij}$ is the amplitude of hopping from site $i$ to site $j$ in a lattice ($\lngl\ph{ij} \rngl$ notation restricts hopping to nearest neighbor sites only) , operator $c\y_{i\sig}$ creates and $c\py_{i\sig}$ destroys an electron with spin $\sig$ at site $i$ respectively ($\n_{\sig}=c\y_{i\sig}c\py_{i\sig}$), $U$ is the strength of on-site local Coulomb interaction, $\eps_d$ is the orbital energy of electrons at each site, and $\mu$ is the chemical potential of the system. The LMA formalism is built up on the fact that the transverse spin-flip scattering can play a crucial role in determining the energy scale that governs the physics of correlated lattice models. Such transverse spin-flip scattering process appears as a polarization propagator in the standard diagrammatic perturbation theory and 
the site-diagonal term can be mathematically written as a convolution integration of `bare' propagators $\mcG_\sig$~\cite{fetter:walecka:polarization}
\blgn
\zpi_{\sbs}\omb=\f{i}{2\pi}\nint d\om'\, \mcG_{\bsig}(\om')\, \mcG_\sig(\om'-\om)\,.
\elgn
Here we consider  $\mcG_\sig$ to be the spin symmetry broken or {\it unrestricted} Hartree Fock (UHF) propagator: $\mcG_\sig\omb=1/(\om-\Sig^0_\sig-\Del\omb)$; where
$\Sig^0_\sig=\eps_d-\mu+U\lngl\n_{\al\sig}\rngl=\eps_d-\mu+\f{U}{2}(n-\sig m)$ is called the 
UHF self-energy, 
$n\equiv\sum_\sig\lngl \n\rngl =-\pim\sum_\sig\nint d\om\,\mcG_\sig\omb$, 
$m\equiv\sum_\sig \sig \lngl n_\sig\rngl=-\pim\sum_\sig\nint d\om\,\sig\,\mcG_\sig\omb$; 
and $\Del\omb$ is the Feenberg self-energy~\cite{economou}.

Similar polarization propagators appear also in the higher order terms of the perturbation series and a careful observation infers that the local (site diagonal) terms of all orders can be arranged in a geometric
progression and hence the net polarization propagator $\tpi_\sbs$ can be expressed as~\cite{logan:eastwood:tusch:jpcm97}
\blgn
\tpi_\sbs\omb=\zpi_\sbs\omb/(1-U\,\zpi_\sbs\omb)\,.
\elgn
$\tpi_\sbs$ is often termed as the  RPA (random phase approximation) polarization propagator.
It leads to a dynamic self-energy contribution that can be expressed in terms of another convolution integral~\cite{logan:eastwood:tusch:jpcm97}:
\blgn
\Sig_\sig\omb=\f{U\sq}{2\pi} \nint d\om' \mcG_\bsig(\om-\om')\, \tpi_\sbs(-\om')
\elgn
Thus collecting the static UHF part $\Sig^0_\sig$ as well, we obtain the total self-energy:
\blgn
\stot_\sig\omb=\Sig^0_\sig+\Sig_\sig\omb\,.
\elgn
The two spin-dependent self-energies give rise to two interacting Green's function $G_\sig\omb=(\mcG_\sig^{-1}\omb-\Sig_\sig\omb)\inv$, which is not directly useful in the case where spin-symmetry is not actually broken (usual paramagnetic case). Therefore to calculate the impurity Green's function in  the DMFT context, we find the spin-averaged Green's function 
\blgn
G\omb=\hf(G_\ua\omb+G_\da\omb)
\label{eq:G:avgd}
\elgn
and obtain a spin-independent self-energy by exploiting the Dyson's Eq. ($G\inv=\mcG\inv-\Sig$):
\blgn
\Sig\omb&=\hf(\stotu\omb+\stotd\omb)\non\\
&\quad+\f{[\hf(\stotu\omb-\stotd\omb)]\sq}
{\mcG\inv\omb-\hf(\stotu\omb+\stotd\omb)}\, 
\label{eq:Sig:cumbersome}
\elgn 
where $\mcG$ is the host Green's function of DMFT's
effective impurity model $\mcG\omb=1/(\om-\Del\omb)$ with $\Del\omb$
playing the role of the hybridization function. Instead of using \eref{eq:Sig:cumbersome},
which apparently looks cumbersome, we find $\Sig$ by writing
\blgn
G\omb=\f{1}{\g\omb-\Del\omb};\; \g\omb\equiv\om+(\mu-\eps_d)-\Sig\omb\,.
\elgn
We similarly can express: $G_\sig\omb=1/(\g_\sig\omb-\Del\omb)$ with $\g_\sig\omb\equiv\om-\stot_\sig\omb$ and then by exploiting  \eref{eq:G:avgd} we determine:
\blgn
\g\omb=\f{2\g_\ua\omb\g_\da\omb-[\g_\ua\omb+\g_\da\omb]\Del\omb}
{\g_\ua\omb+\g_\da\omb-2\Del\omb}\,.
\label{eq:gamma:gup:gdn}
\elgn
To attain the DMFT self-consistency on the lattice side, we  
find the local Green's function by performing the Hilbert transform: 
$G\omb=\nint d\eps\, D_0(\eps)/(\g\omb-\eps)$
for a given non-interacting lattice density of states (DoS) $D_0\omb$. 
Furthermore, for the metallic phase, in order to ensure the Fermi-liquid property,
we add the following \emph{symmetry restoration} condition (pinning of the 
spectral density to the non-interacting limit at the Fermi level) into the DMFT equations:
\blgn
\sum_\sig\, \sig\Sig_\sig(0)=|m|\,U\,.
\label{eq:SR}
\elgn  

In the p-h symmetric case, we use the condition $\eps_d-\mu=-U/2$, where $\mu$ is chosen
to be zero in practice. However, for the asymmetric case, we do not have such a simple relation between the orbital energy and the Coulomb interaction strength.  Also there should be a shift $\del\mu$ from the chemical potential $\mu$, which is set to be zero in the symmetric case. Therefore UHF Green's function gets modified as $\mcG_\sig\omb=1/[\omp-\teps+\sig |m|U/2-\Del\omb]$ where  
$\teps\equiv\eps-\del\mu\equiv\eps_d-\mu+Un/2-\del\mu$
which is zero only in the half-filled case ($\eps=0$, $\del\mu=0$).

Now there are two important algorithmic remarks that we would like to  make here:
\begin{enumerate}[(i)]
\item We parametrize a quantity $x\equiv\hf|m|U$ and for a given $x$ 
we determine $U$ by the symmetry restoration condition (\eref{eq:SR}). This step is common
to both the half-filled and the away from half-filled cases. Note that for the insulating
case, this condition is not required, however, a pole arises at $\om=$ in $\im\tpi_\sbs\omb$,
which needs to be taken care by analytically adding its weight to the self-energy~\cite{logan:eastwood:tusch:jpcm97}.
\item Once we find $U$,  we calculate $\Sig\omb$ and $G\omb$
for a fixed $\teps=\eps_d-\mu+Un/2-\del\mu$,
then, setting $\mu=0$, we find the $\eps_d$ by self-consistently satisfying
Luttinger's sum-rule~\cite{luttinger:theorem:pr60,langer:ambegaokar:pr61,muller-hartmann:zpb89}: $\int_{-\infty}^\mu d\om\,G\omb \f{\pd \Sig\omb}{\pd \om}=0$ . 


An asymmetry parameter $\eta\equiv1+2\eps_d/U$ ($\mu=0$)
is introduced to quantify  p-h asymmetry in our calculations.
Note that for the symmetric case,
$\eps_d=-U/2$ and hence $\eta=0$.
 
\end{enumerate}

\section{Results and discussions}
We separate our results and corresponding discussions into A:
the particle-hole symmetric or half-filling ($n=1$) case and B: the case away from 
it ($n\ne 1$). Our discussions mostly comprise of the properties of single-particle spectral density
and analysis following that at different parameter regimes at zero temperature. 
Note that as a part of DMFT method, the hopping amplitude in \eref{eq:HM} is taken to be uniform and we define a new hopping amplitude $\ts$ such that $t_{ij}= \ts/\sqrt{z}$,
 $z$ being the coordination number. Throughout the paper we choose $\ts=1$ for our calculation, and results are obtained for the $d$-dimensional hypercubic lattice ($z=2d$) though similar features are tested
in the Bethe lattice as well. The non-interacting DoS 
of the lattice is defined as $D_0\omb\equiv 1/(\sqrt{\pi}\ts)\exp(-\om\sq/\ts\sq)$. At the end of Section B, we keep a special subsection for the optical properties, for which we use the standard Kubo formula from the linear response theory~\cite{pruschke:cox:jarrell:prb93,hbar:nsv:ijmpb11}.
 
\subsection{A. At half-filling} 
\subsubsection{Universal scaling behavior of spectral density}
\begin{figure}[!htp]
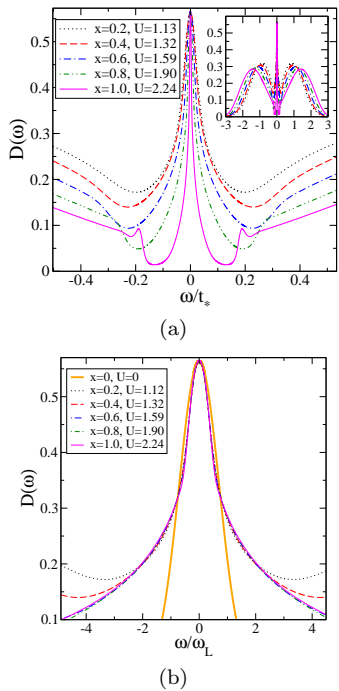

\subfigure[]{
\includegraphics[height=4cm,clip]{FIGS/Spf_all_sym_ZT.eps}
\label{fig:spf:scaling:ZT:a}
}
\subfigure[]{
\includegraphics[height=4cm,clip]{FIGS/Scaled_Spf_sym_all.eps}
\label{fig:spf:scaling:ZT:b}
}
\caption{Spectral densities and their scaling collapse. (a) Abrikosov-Suhl resonance appears at the Fermi level ($\om=0$) and the resonance width decreases with increasing $x$ or $U$. Inset: full spectra for the same. (b) Scaling collapse of
spectral densities when the frequency axis is scaled by the low-energy scale $\om_L=Z\ts$. Note that the collapse deviates from the non-interacting curve ($U=0$) almost immediately away from the Fermi level.}
\label{fig:spf:scaling:ZT}
\end{figure}

The key investigative question that arises at the half-filling case is 
whether the Mott transition is seen at large Coulomb interaction $U$, which should be 
reflected by formation of a gap at the Fermi level (set as $\om=0$ in our convention) in the spectral density, $D\omb=-\pim G\omb$. 
Before we seek an answer, we first look at the low-energy behavior of the the spectral density for small interaction strength $U$ and hence for small $x$. \Fref{fig:spf:scaling:ZT:a} ~shows the presence of finite DoS
at the Fermi level in the form of quasiparticle or the Abrikosov-Suhl resonance,
clearly signaling a metallic phase (The inset figure shows the usual three-peak full spectral density at various $U/\ts$'s.). As we may expect from the construction of our methodology, all the resonance peaks are pinned at the non-interacting value at the Fermi level: $D(0)=D_0(0)=1/\sqrt{\pi}$. This is known as the Luttinger pinning~\cite{vollhardt:etal:jpsj05}, which is a direct consequence
of the Luttinger's sum-rule mentioned in the earlier section~\cite{muller-hartmann:zpb89}. The resonance width shrinks gradually as we increase $x$ or $U$, and we can associate an effective low-energy scale,
$\omega_L=Zt_*$, determined from the quasiparticle residue $Z=1/(1-\pd_\om\re\Sig\omb|_{\om=0})$, proportional to the width of the resonance.
%
%
From \fref{fig:spf:scaling:ZT:b} we can see that all spectral densities collapse to 
a universal value around the Fermi level when we scale the frequency axis by $\om_L$. 
Nevertheless the collapsed spectral density seems to deviate from the non-interacting 
limit almost immediately away from the Fermi level.
Thus, even though adiabatic continuity at the Fermi level is maintained in our formalism
through \emph{symmetry restoration} in \eref{eq:SR}, the renormalized non-interacting limit (RNIL) description is seen to be invalid. This can be explained if we look at the self-energy behavior at low-frequency.
%
\begin{figure}[!htp]
\includegraphics[height=5cm,clip]{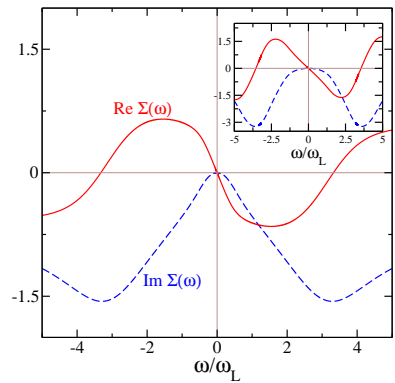}
\caption[{Imaginary and real part of the self-energy at
$T=0$ (LMA and IPT).}]
{Reason for non-collapse with the non-interacting DoS. Dashed lines (blue in color) and full lines (red in color) are the imaginary and real part of the
self-energy respectively. The static part $\Sig(0)=U/2$ has been subtracted from the real part. $\im\Sig\omb$ grows far more rapidly from $\om=0$ in LMA (main panel)
than that in IPT (inset) suggesting that not only in the strong coupling regime, but also in the intermediate correlation regime, incoherent scattering effects become
important at energies even slightly away from the Fermi level . The interaction strength for the LMA: $U=1.13\ts\; (x=0.2\ts)$ and for the IPT: $U=3.0\ts$.}
\label{fig:im:re:Sig}
\end{figure}    

The RNIL assumes that contribution from $\im\Sig\omb$ is negligible compared to the contribution from $\re\Sig\omb$ at low $\om$ since the former vanishes as $\om\to 0$
with one power of $\om$ ($\propto \om\sq$) higher than the latter ($\propto \om)$. This assumption does hold in IPT over a large
interval around the Fermi level. However, the contributions from both real and imaginary part of $\Sig\omb$ become comparable when the coefficient
of imaginary part becomes large enough. \Fref{fig:im:re:Sig} shows that the slope change in $\im\Sig\omb$ away from $\om=0$ is faster in LMA
(shown in the main panel) compared to that in IPT (shown in the inset). This signifies that 
the incoherent scattering commences immediately after the Fermi level once we incorporate 
transverse spin-flip mechanism into the diagrammatic perturbation theory.
\subsubsection{Emergence of low energy scale in susceptibility}
Since spin-flip scattering is responsible for the rise of  the Kondo energy scale of an impurity model and the impurity physics persists in a lattice through the self-consistency of the DMFT formalism, it is natural to intuit such a scale in LMA.
Moreover, in the strong-coupling Kondo regime (Fermi liquid) the Kondo scale 
should be proportional to $\om_L=Z\ts$~\cite{hewson}. 
%
\begin{figure}[!htp]
\includegraphics[height=5cm]{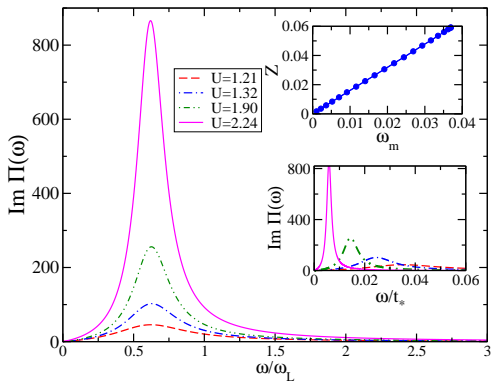}
\caption{Bottom inset: The peak position of $\im\Pi\omb$ gives rise to an energy scale $\om_m$ for various $U$'s. Main panel: The position become universal when the frequency is scaled by $\om_L=Z\ts$ signifying a proportional relation between $Z$ and $\om_m$. Top inset: $Z$ vs $\om_m$ plot is a nice straight line with slope 1.59.}
\label{fig:Pi:n:wm}
\end{figure}
%
The bottom inset of \fref{fig:Pi:n:wm} shows that $\im\Pi\omb$ has a maximum or peak at $\om=\om_m$. Once we scale the frequency axis by $\om_L$ (main panel of \fref{fig:Pi:n:wm}), the positions of those peaks 
fall at the same value, which clearly indicates a proportional relation between $\om_m$ 
and the Fermi liquid scale $\om_L$. Presence of such maxima gives rise to maxima in response functions such as imaginary part of spin susceptibility and absorption spectrum (real part of dynamic conductivity). Recent DMFT study using various impurity solvers have shown that indeed position of the imaginary part of local spin susceptibility become universal while frequency axis is scaled by $\om_L$~\cite{grete:etal:prb11}. 
The top inset of \fref{fig:Pi:n:wm} reaffirms our statement showing a linear dependence of $Z$ on $\om_m$ with a slope 1.59 (proportionality constant).
\subsubsection*{Mott transition and presence of hysteresis}
\begin{figure}[!htp]
\includegraphics[height=5cm,clip=] {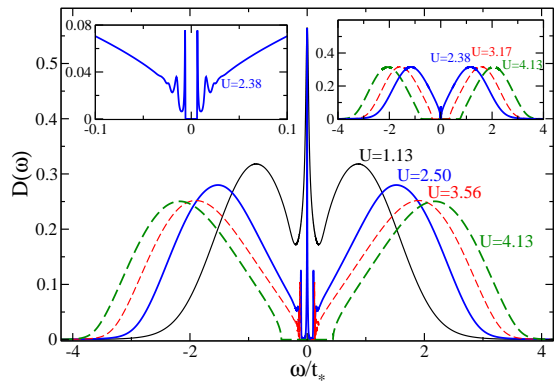}
\caption[{Mott transition}]
{Mott transition reflected from the spectral density evolution as $U$ changes at zero temperature (for the hypercubic lattice). Main panel: quasiparticle resonance shrinks 
as $U$ is increased. At $U=3.56\ts$, a clear gap opens up at the Fermi level signaling 
the Mott metal-to-insulator transition. The gap gets enhanced 
as $U$ is increased further. Right panel: Starting from a Mott insulator if $U$ is decreased,
the gap at the Fermi level decreases and eventually closes at $U<2.38\ts$. Left panel: 
A finite gap still persists at $U=2.38\ts$. Therefore an extrapolation method is 
required to find the critical value $U_{c1}$ where insulator-to-metal transition happens 
(see \fref{fig:gap:Z:Vs:U:ZT}).}
\label{fig:Spf:MIT:IMT}
\end{figure}

\begin{figure}[!htp]
\centering
\includegraphics[height=5cm,clip]
{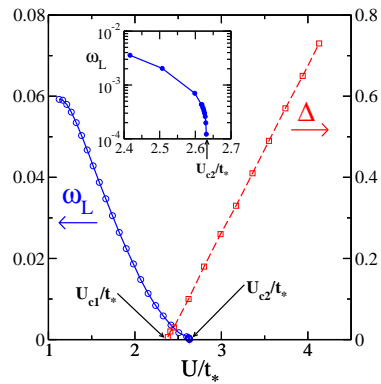}
 \caption[{Vanishing of Fermi liquid
scale signifying Mott transition}]
{Decay of the low-energy scale $\om_L=Z\ts$
with increasing $U/\ts$ (solid line).
Inset shows the same in log-scale. $\om_L$
seems to vanish at $U_{c2}=2.64\ts$. Note that we have been able to reach a value of
the low energy scale $\sim 10^{-4}t_*$, which requires very high precision calculations.
The dashed line shows that the spectral gap $\Delta$
of the Mott insulator decreases linearly
with decreasing $U/\ts$ and closes
at $U_{c1}=2.36\ts$. 
}
\label{fig:gap:Z:Vs:U:ZT}
\end{figure}

It is already mentioned that 
the width of the quasiparticle resonance shrinks gradually
as $U/\ts$ is increased, which disappears finally by opening
up a gap at the Fermi level. Thus our primary question is answered
and indeed an interaction-driven
metal-to-insulator transition, i.e. Mott transition
occurs when interaction strength is greater than 
a critical value, i.e. $U\ge U_{c2}$.  
For the hypercubic lattice (HCL), we find approximately $x_{c2}=1.3\ts$ which implies
$U_{c2}\simeq 2.8\ts$. In the main panel of \fref{fig:Spf:MIT:IMT} we see that a gap opens 
at the Fermi level in the spectral density for $U=3.56\ts$.
%
%
%
The estimation of $U_{c2}$ is carried out through an extrapolation of the zero crossing of the 
low energy scale $\om_L$  with increasing $U$ (see line with open circles in \fref{fig:gap:Z:Vs:U:ZT}).
In IPT, it has been seen~\cite{hbar:nsv:ijmpb11}, in the zero temperature evolution
of spectral densities with interaction strength, that
there exist two transition points $U_{c1}$ and $U_{c2}$
depending on whether we are
changing $U$ from the metallic or insulating side.
Therefore it is natural to ask: 
If we start from an insulating regime
and keep on decreasing $x$ (hence $U$), do we get an insulator to
metal transition at the same point that we have mentioned
above? The right inset of \fref{fig:Spf:MIT:IMT}
shows that we find that the gap decreases as we decrease $x$ from $2.0\ts$
($U=4.13\ts$) and it appears that the gap closes
at $x\sim 1.06\ts$, i.e. $U=2.38\ts$ 
%
%
However, the gap is
truly not zero at $U=2.38\ts$ as the left inset of \fref{fig:Spf:MIT:IMT}
shows in a zoomed view.
For this reason we plot the gap ($\Del$) as a function of $U/\ts$
in \fref{fig:gap:Z:Vs:U:ZT} (dashed line with open squares).
We find that $\Del$ almost linearly decreases with
$U/\ts$ and to our estimation $U_{c1}\simeq 2.36\ts$.
Thus similar to the IPT result, LMA also
shows presence of a coexistence regime
(possibility of having both metallic and insulating solutions)
and hence hysteresis driven by interaction.
The width of the coexistence regime, i.e.
$\Del U_c\equiv U_{c2}-U_{c1}$ is $0.44\ts$ for the HCL which is a little less
than that found in IPT ($\Del U_c\sim 0.7\ts$)~\cite{hbar:nsv:ijmpb11}. 
However, in Bethe lattice the coexistence regime is further small~\cite{eastwood:phd:98}.
%

\subsubsection{Spectral moment sum-rules}
\begin{figure}
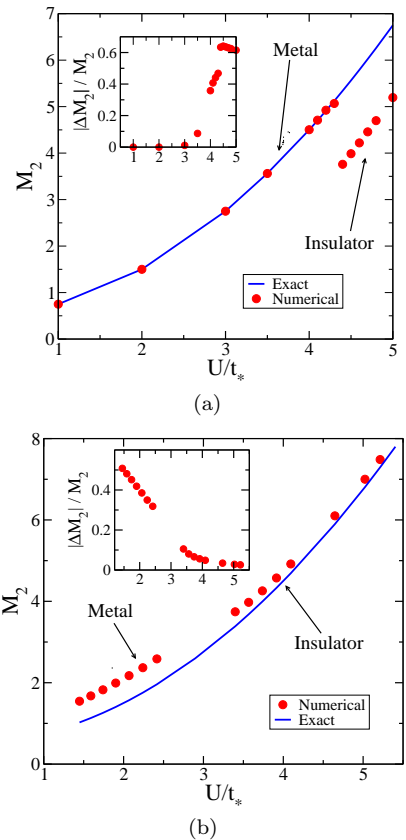

\subfigure[]{
\includegraphics[height=5cm,clip]{FIGS/M2_n_ErrM2_vs_U_IPT.eps}
\label{fig:M2:err:IPT}
}
\subfigure[]{
\includegraphics[height=5cm,clip]{FIGS/M2_n_ErrM2_vs_U_LMA.eps}
\label{fig:M2:err:LMA}
}
\label{fig:M2:err}
\caption{Second spectral moment $M_2$ in (a) IPT and (b) LMA for various 
interaction strengths. The insets show the respective absolute value of relative errors.}
\end{figure}

Spectral moments are often considered to be important in testing the robustness of a certain
numerical or analytical method for a many-body problem~\cite{white:specmom:prb91,turkowski:freericks:prb06}. A $m$-th
spectral moment is defined as $M_m\equiv\nint d\om\, \om^m D\omb$. $M_0=1$ is true for any model and for the Hubbard model one can find: $M_1=\eps_d-\mu+U\lngl n_\sig\rngl$, $M_2=\sk\eps_\kv^2+(\eps_d-\mu)^2+U\lngl n_\sig\rngl[2(\eps_d-\mu)+U]$, $M_3=M_1 M_2$. Here $\eps_\kv$ is the dispersion of the given lattice whose
momentum ($\kv$) sum is nothing but the second spectral moment of the non-interacting DoS, i.e.
$\sk \eps^2_\kv= M_2^0\equiv\nint d\om\,\om^2 D_0(\om)$. For instance, for a Bethe lattice DoS,      
  $D_0(\om)\equiv\f{1}{2\pi\ts\sq}\sqrt{4\ts^2-\om^2}$, $M_2^0=\ts\sq$, and for our HCL DoS, 
$M_2^0=\hf\ts\sq$. For half-filling case, we obtain further simplification: 
$M_1=0$, $M_2=M_2^0+U^2/4$, $M_3=0$. Now in our IPT calculation, $M_0=1$, $M_1$ and $M_3$ varies within the order $10^{-3}$-$10^{-5}$ and $10^{-2}$-$10^{-3}$. On the other hand,
in LMA, the errors goes to the order of $10^{-10}$-$10^{-12}$ as one approaches towards 
higher $U$. In the half-filling case, specifically the second moment $M_2$ becomes very crucial. \fref{fig:M2:err:IPT} shows that numerically calculated $M_2$  significantly agrees with the expected analytical value, however the agreement severely breaks down in
the insulating regime ($U>U_{c2}=4.4\ts$). On the contrary, in LMA, the agreement is comparatively poor in the metallic side, but the difference ($\Del M_2$) between the exact and numerical values decreases as $U$ increases and it appears that $\Del M_2\to 0$ as $U\to\infty$ (see \fref{fig:M2:err:LMA}). The absolute values of the relative errors are shown in the insets. A very recent paper~\cite{lu:hoppner:gunnarson:haverkort:prb14} has reported higher 
accuracy in the spectral moments up to the third order using a new alternative diagonalization-based QIS. Nevertheless the method is heavily expensive in computation time and limited by finite number of sites and errors could be introduced by the broadening over discretization, which fails to ensure the Luttinger pinning at the Fermi level. 
%
%

\subsubsection{Strong correlation universality}
\begin{figure}
\includegraphics[height=5cm,clip]{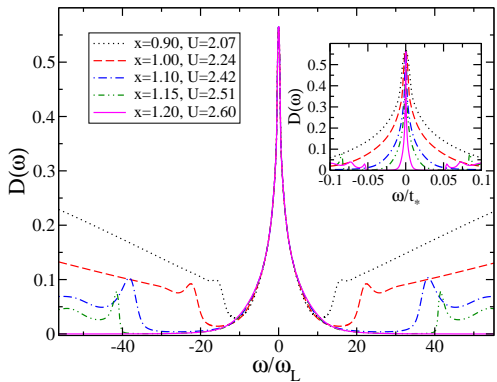}
\caption{Scaling universality at strong correlation strength: as $U\to U_{c2}^{-}$, the collapse
of the spectral density extends all the way after the frequency being scaled by $\om_L$, despite being limited to the low energy scale regime, which could be order of  $\om_L$. For example, 
the scaling agreement between spectral densities at $U=2.51\ts$ and $U=2.60\ts$ runs up to
$\sim 30\om_L$. The inset shows the same spectral densities without the scaling.} 
\label{fig:Spf:sc:universality}
\end{figure}

As noticed in \fref{fig:spf:scaling:ZT:b}, the spectral density 
seems to assume a universal form $D\omb=D(\om/\om_L)$ leading to collapse
of $D\omb$ up to a certain frequency range. As $U/\ts$ increases, this range
keeps on increasing and 
close to the Mott transition, we find scaling collapse 
in the spectral densities for decades of $\om_L$
(see main panel of \fref{fig:Spf:sc:universality}):
$U$ ranging from $2.07\ts$ to $2.60\ts$),
when the frequency axis is scaled by the same energy scale. Moreover, this universal regime extends to higher and higher frequencies as we increase $U/\ts$
suggesting that in the limit $U\rightarrow U_{c2}^-$, the universal scaling region extends to all the way till the frequency reaches one of the Hubbard bands.
%
%
The universal scaling form is seen to be very different from the RNIL suggesting very non-trivial tails of the spectral function for large $\om/\om_L$.
These tails should manifest themselves in transport and other finite temperature/frequency
properties that would be an interesting feature to look for in experiments~\cite{nsv:victoria:logan:jpcm03}.
%

\subsection{B. Away from half-filling}   
%
\subsubsection{Spectral density: empty orbital, mixed valence, 
and doubly occupied orbital states}  

\begin{figure}[!htp]
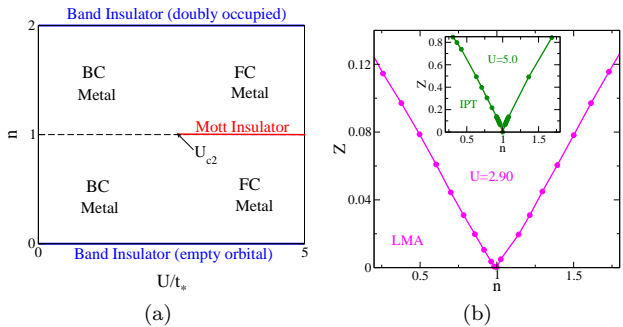

\subfigure[]{
\centering\includegraphics[height=3.8cm,clip]{FIGS/PD_n_Vs_U.eps}
\label{fig:PD:n:Vs:U}
}
\subfigure[]{
\includegraphics[height=3.5cm,clip]{FIGS/Z_Vs_n_LMA_IPT.eps}
\label{Z:vanish}
}
\caption{(a) Phase diagram on the occupancy-interaction ($n$-$U$) plane. 
 The region is bounded along the filling axis by empty orbital and doubly occupied band insulator lines. The metal emerging by doping a Mott insulator side is known as the filling controlled (FC) metal ($U>U_{c2}$) and remaining region is the band-width controlled (BC) metal since interaction is low. In case of LMA, $U_{c2}=2.8 \ts$.
(b) Disappearance of quasiparticle residue $Z$ as occupancy approaches the half-filling
value $n=1$ in
IPT (Inset: $U=5.0\ts$) and LMA (Main: $U=2.9\ts$).
}
\end{figure}

Before we embark on the results, we first make a few qualitative remarks. When the electron density is not equal to one per site, i.e. away from the half-filling, there are always empty sites available for electrons/holes to hop without
encountering the Coulomb repulsion $U$. Therefore we can get Mott insulators only when the filling reaches the half-filled value ($n=1$). However, there can be special situations, namely $n=2$ where electron's hopping is forbidden since orbitals at all sites are fully
(doubly) occupied. This leads to an insulator, which is in fact a band insulator. Similarly for $n\to 0$ case, there will be only a few electrons left for conduction, or from the hole point of view, the sites will be fully occupied again and will lead to a band insulator. Thus at zero temperature we can divide the $n$-space into five distinct regimes, viz.
(i) empty orbital ($n\to 0$), (ii) mixed valence-I ($0<n<1$), (iii) symmetric metal or Mott insulator ($n=1$), (iv) mixed valence-II ($1<n<2$), and (v) doubly occupied orbital ($n\to 2$).
The regimes (iv) and (v) are p-h symmetric counterparts of (ii) and (i) respectively.
%
\begin{figure}[!htp]
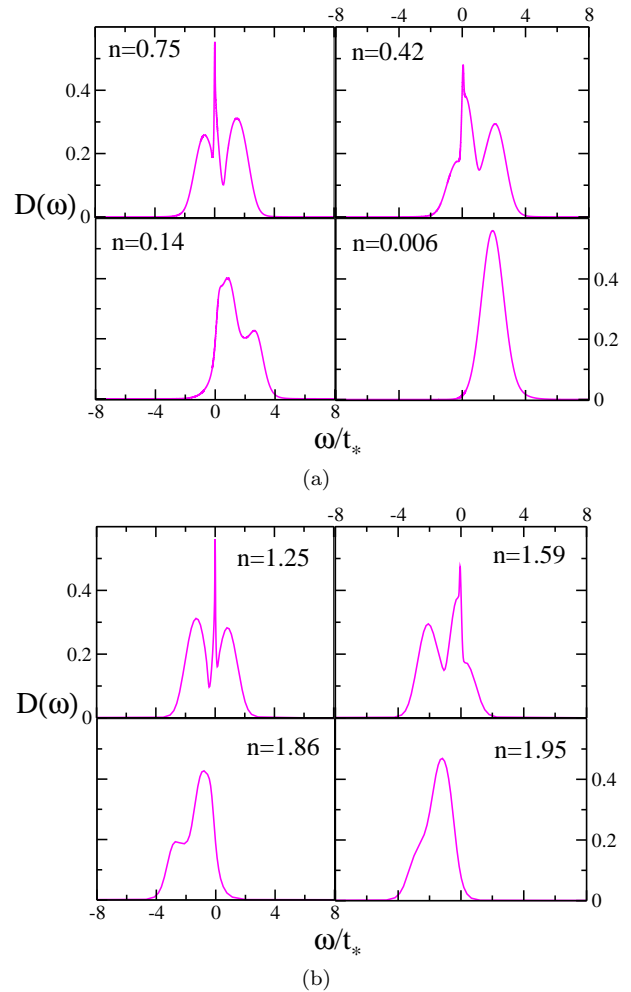

\subfigure[]{
\includegraphics[height=6cm,clip]{FIGS/Spf_n_decreasing.eps}
\label{fig:spf:n:lt:1}
}
\subfigure[]{
\includegraphics[height=6cm,clip]{FIGS/Spf_n_increasing.eps}
\label{fig:spf:n:gt:1}
}
\caption[{Spectral density evolution with filling from $n<1$ to $n\simeq 0$ (LMA)}]
{Evolution of spectral densities for $x=0.5\ts$ ($U\simeq 1.5\ts$) at $n\ne 1$ (away from half-filling), 
for the hypercubic lattice (HCL). Parent (half-filled, i.e $n=1$) phase is metallic. 
(a) Evolution from $n\le 1$ to $n\to 0$.
(b) Evolution from $n\ge 1$ to $n\to 2$. }
\label{fig:spf:asym}
\end{figure}
%
\fref{fig:spf:n:lt:1} and \fref{fig:spf:n:gt:1} show the evolution
of spectral density towards the two extremes ( regime (i) and regime (v) ) for the hypercubic lattice, starting from a half-filled Fermi liquid metal ($n=1$).
In the first case, the lower Hubbard band starts moving towards the Fermi level ($n=0.75$) with decreasing its height compared to the upper Hubbard band, then
it coalesces with the quasiparticle resonance ($n=0.42$) where resonance itself shifts away from the Fermi level.
Gradually the lower Hubbard band and the qausiparticle features do not remain
significant any more ($n=0.14$) and the density just behaves like a non-interacting
one, situated above the Fermi level, thus being 
a band insulator with the band edge at the Fermi level. Similarly in the second case, the upper Hubbard band moves towards the Fermi level and finally the lower Hubbard band occupies the whole spectral region and the system becomes
an empty orbital band insulator (regime (i)). Thus \fref{fig:spf:n:lt:1} and \fref{fig:spf:n:gt:1} reflect the the fact that 
for a particle with $1\le n\le 2$ has its hole counter-part in
$0\le n\le 1$. A schematic phase diagram on the occupancy-interaction plane at zero temperature is shown in \fref{fig:PD:n:Vs:U}.
%
%
The filling control MIT can be inferred by looking at quasiparticle residue $Z$, as it continuously vanishes at the half-filling ($n=1$). \fref{Z:vanish} shows this behavior
for both LMA (main) and IPT (inset).
\subsubsection{Pseudogap formation and strong-coupling universality}
\begin{figure}[!htp]
\centering\includegraphics[clip,height=5cm]{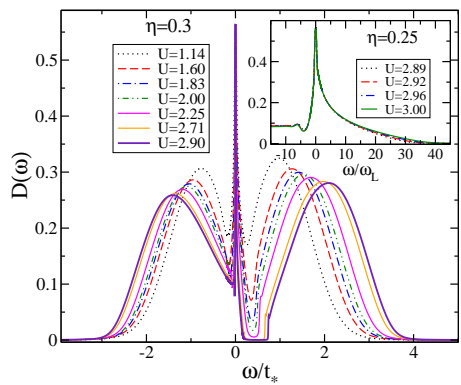}
\caption
{Main: Spectral densities at various $U/\ts$'s at asymmetry parameter
$\eta=0.3$. A pseudogap (a gap close to the Fermi level) 
forms in the spectral density at $U= 2.9\ts$.
Inset: Scaling
universality at strong interaction values at $\eta=0.25$: 
$U/\ts=$ 2.89, 2.92, 2.96, and 3.00. The
universal region extends to very large values of
$\omega/\omega_L$ and the universal scaling form
is seen to be very different from the
renormalized non-interacting Gaussian form.} 
\label{fig:Spf:asym:pseudogap:sc:scaling}
\end{figure}

The main panel of \fref{fig:Spf:asym:pseudogap:sc:scaling}  shows that after certain 
$U/\ts$ ($\sim 2.7 \ts$)
%
a pseudogap starts to form near the Fermi level
(\emph{pseudo-} since the gap does not open exactly at the Fermi level).
The gap increases as we increase $U/\ts$ further.
We have noticed that pseudogap has the same width
as the gap in the Mott insulator has in
half-filling. It seems that the quasiparticle
weight never vanishes at any large finite $U/\ts$
above $U_{c2}/\ts$ and hence the pseudogap never
touches (however close it may be) the Fermi level. 
This is expected because once we go away from half-filling, even by
infinitesimal doping, we never expect a Mott
transition. The pseudogap feature, however, is not observed using 
IPT~\cite{kajueter:Kotliar:moeller:prb96}. Therefore, the feature
might be tied to the transverse spin-flip scattering process inherent in LMA. 

%

Similar to the half-filled case, the scaling universality for strong 
interaction strength extends to very large
frequencies beyond the low-energy Fermi liquid
scale $\om_L$ (see inset of \fref{fig:Spf:asym:pseudogap:sc:scaling}  ) and it appears that as we increase $U/\ts$ further, the scaling agreement extends further and at strong-coupling limit
 ($U\to\infty$), we expect the scaling universality will extend all the way in frequency, 
 $\om\to \pm \infty$, since the Hubbard bands are positioned at $\pm\infty$ now.
\subsubsection{Optical conductivity}
\begin{figure}[!hbp]
\includegraphics[clip,height=5cm]{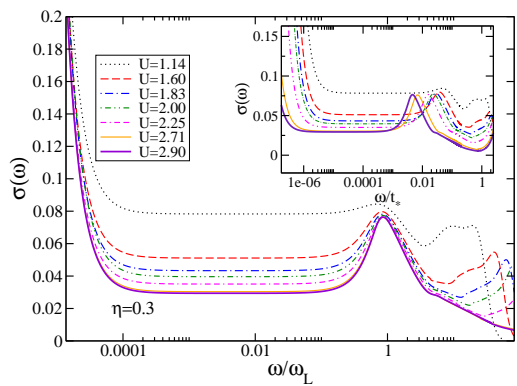}
\caption
{Scaling behavior in the optical conductivity $\sig\omb$ for the asymmetry parameter $\eta=0.3$ and different interaction strengths. The first absorption peaks arise at
$\om=\om_L$ (Main panel), as revealed from the scaling of the frequency axis by $\om_L$,
leaving all peaks appear at $\om/\om_L=1$ (Inset).}
\label{fig:scaling:OC:asym}
\end{figure}

\begin{figure}[!htp]
\includegraphics[clip,height=5cm]{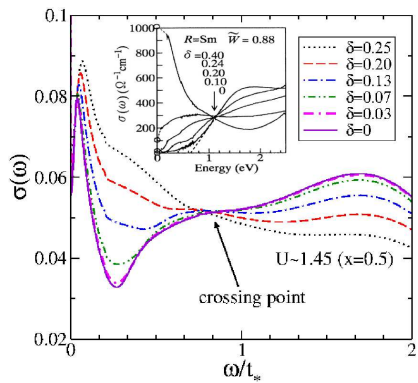}
\label{fig:OC:asym:isosbestic:expt}
\caption[{LMA results for optical conductivity with various hole concentrations and comparison with experiments}]{Main: LMA results for optical conductivity with various hole concentrations: $\del=1-n$. Inset: Optical conductivity for Sm$_{1-x}$Ca$_x$TiO$_3$ where $\tilde W$ is the bandwidth done by a tight-binding calculation, mentioned in Ref.~\citenum{katsufuji:prl95}, normalized to that
of LaTiO$_3$.}
\label{fig:OC:asym:isosbestic}
\end{figure}

With an ambition to derive some physical properties out of our zero temperature
spectral densities, we seek the optical properties. Being doped and hence metallic in nature, a divergent Drud\'e peak appears at $\om=0$ of the optical conductivity $\sig\omb$, accompanied by an absorption peak positioned at $\om=\om_L$ (see inset of \fref{fig:scaling:OC:asym}, Drud\'e peaks are out of scale). This uniqueness of the peak position becomes evident when we divide the frequency axis by corresponding $\om_L$ for various $U/\ts$ at fixed asymmetry parameter $\eta$. For instance,
at $\eta=0.3$ all of the first absorption peak $\sig\omb$ for different $U/\ts$, arise at 
$\om/\om_L=1$ (see main panel of \fref{fig:scaling:OC:asym}). This result is very significant
because any experimental probe that finds the absorption spectra of a material
in a certain condition, can easily determine the associated low-energy scale of it by 
looking at the position of the first absorption peak. 
Moreover, this universal feature of the absorption peaks implies that such a 
universality is merely a signature of a Fermi liquid and does not get affected by doping as long as the phase remains a Fermi liquid. 
%
%
%
Another interesting feature is noticed when the optical
conductivity is computed for different hole dopings $\del\equiv 1-n$, 
keeping the interaction unchanged. The main panel of \fref{fig:OC:asym:isosbestic} depicts $\sig\omb$ at various dopings ($\del$ ranging from 0 to 0.25) where we notice that
a universal crossing point appear around $\om\simeq\ts$.  
This behavior does indeed bear close similarity to
the experiments on compounds of the formulae 
R$_{1-x}$Ca$_x$TiO$_{3+y}$, R representing rare-earth metals, done by Katsufuji \etal~\cite{katsufuji:prl95} (see inset of \fref{fig:OC:asym:isosbestic:expt}). Similar spectral weight transfer through a universal point or a point-like region in the cuprates (e.g. La$_{2-x}$Sr$_x$CuO$_x$~\cite{uchida:prb91} and  Pr$_{2-x}$Ce$_x$CuO$_4$~\cite{arima:prb93}), Sr doped LaCoO$_3$~\cite{tokura:prb98}, and very recently in NiS$_{2-x}$Se$_x$~\cite{perucchi:NiSe} has been observed. Such a universal point is termed as the \emph{isosbestic point} and presence of it is often considered
to be reminiscence of correlation effect~\cite{eckstein:kollar:vollhardt:jltp07,greger:kollar:vollhardt:prb13}.
%
%
%

\section{Summary}
In summary, we must say that, within the LMA+DMFT framework, our work seeks the unexplored part of the single orbital Hubbard model  
i.e. the metallic phase at arbitrary filling and the phase
diagram on the filling-interaction plane at zero temperature. LMA shows Mott metal-to-insulator
transition like many other solvers of the DMFT impurity problem. However, the transition 
point differs from many other methods, mainly due to lower value of the quasiparticle residue
(at least by one order of magnitude, see \fref{Z:vanish} for instance). This is also what prevents LMA spectral densities to be benchmarked with that from other numerical methods. If we leave this issue aside for a moment,
we can see LMA successfully captures all essential physics of the Hubbard model. For instance,
the spectral density with three-peak structure (quasiparticle resonance plus two Hubbard bands)
in the metallic phase. Specifically the Luttinger pinning of the spectral density is excellently obeyed in LMA, which is a difficult challenge for many other numerical methods.
Being semi-analytical, IPT and LMA both possess similar advantages, e.g. being computationally
non-expensive and capable of producing qualitatively correct physics. However, the spectral moment sum-rule breaks down for IPT in the insulating regime, where LMA plays its best role. 
The strong-coupling universality and presence of pseudogap may require deeper understanding
and connection to the impurity model physics~\cite{nsv:logan:epjb04}. The optical properties also reflect universal
features and a finite temperature extension to it, which in principle requires no extra formalism, could be a topic of our follow-up paper, which may attempt to find some answers to the long-lasting puzzles in experiments of doped Mott insulators~\cite{lee:nagaosa:wen:rmp06}.

\section{Acknowledgments}
For being introduced to the project the author is indebted to N. S. Vidhyadhiraja, whose expertise in LMA+DMFT method offered substantial help. He also thanks DST and DAE of Govt. of India for providing financial support
and scientific resources and Vikram Tripathi for his needful advices. 

\bibliographystyle{apsrev4-1}
\bibliography{refs}

\end{document}